%Paper: hep-lat/9309012
%From: Anil Trivedi <trivedi@yukawa.uchicago.edu>
%Date: Sat, 18 Sep 93 04:36:22 CDT
%Date (revised): Mon, 4 Oct 93 05:18:52 CDT
%Date (revised): Mon, 4 Oct 93 05:27:08 CDT

%&plain
% THE NIELSEN-NINOMIYA ``NO-GO'' THEOREM IS FALSE
% by Anil K. Trivedi; hep-lat/9309012
% TeXsis, or PLAIN TeX (with "mtexsis.tex"); 9 pages

\input mtexsis

\newif\ifLocalProcessing\newif\ifForceTenPointFonts	 %\LocalProcessingtrue
\def\ATunlock{\catcode`@=11} \def\ATlock{\catcode`@=12} \ATunlock
\def\l@m{>\space} \def\t@m{....} \def\sp@ce{\space\space\space}
\immediate\write16{} \immediate\write16{ \l@m Testing for the TeX dialect\t@m}

%---Test for LaTeX---*
%
\ifx\@latexerr\und@fin@d\relax				% sigh of relief
\else							% oh well...
  \newlinechar=`\^^J
  \immediate\write16{%1%
    \sp@ce What?? LaTeX!! LaTeX?!?!?!^^J
    ^^J
    \l@m You cannot print this document with LaTeX.^^J
    \l@m ^^J
    \l@m This preprint uses the TeXsis macro package. If TeXsis is installed^^J
    \l@m on your system, please use the command "texsis", not "latex".^^J
    \l@m ^^J
    \l@m You can also print it with Plain TeX. For that, you need the macro^^J
    \l@m file "mtexsis.tex". If you have this file, please run the command^^J
    \l@m "tex", not "latex"; if you do not have it, please get it from^^J
    \l@m hep-th, hep-lat, etc.; it is loaded automatically if you use Plain,^^J
    \l@m you need NOT edit the file \jobname.tex to add
         "\noexpand\input mtexsis".^^J
    \l@m ^^J
    \l@m If your printer is Postscript, I can also send you a Postscript^^J
    \l@m file of this preprint.^^J
    \l@m ^^J
    \l@m --Anil Trivedi (trivedi@yukawa.uchicago.edu)
    ^^J
                    }%1%
\@@end\fi

%---Test for TeXsis---*
%
\newread\@mtxfile
\ifx\texsis\und@fin@d					% if not TeXsis
  \newlinechar=`\^^J
  \immediate\write16{%1%
    \sp@ce Not TeXsis.^^J^^J
    \l@m Looking for "mtexsis.tex"\t@m
                    }%1%
  \immediate\openin\@mtxfile=mtexsis\relax		% look for mtexsis
  \ifeof\@mtxfile\closein\@mtxfile			% if not found
   \immediate\write16{\sp@ce Not found!!!!}
   \immediate\write16{%2%				% warn...
    ^^J
    \l@m This preprint uses the TeXsis macro package. To print it with^^J
    \l@m Plain TeX, you NEED the file mtexsis.tex (available from hep-th,^^J
    \l@m hep-lat, etc.). It is loaded automatically if you use Plain TeX,^^J
    \l@m you need NOT edit the file \jobname.tex to add
         "\noexpand\input mtexsis".^^J
    \l@m ^^J
    \l@m Of course, IF TEXSIS IS INSTALLED on your system, there is no^^J
    \l@m need for mtexsis.tex: just use the command "texsis", not "tex".^^J
    \l@m OTHERWISE, please get mtexsis.tex and run "tex \jobname" again.^^J
    \l@m ^^J
    \l@m If your printer is Postscript, I can also send you a Postscript^^J
    \l@m file of this preprint.^^J
    \l@m ^^J
    \l@m --Anil Trivedi (trivedi@yukawa.uchicago.edu)
    ^^J
                     }%2%
    \end						% ...and quit!
  \else\immediate\closein\@mtxfile			% if found, load it..
 %
 % However: \texsis in mtexsis.tex calls \twelvepoint. If a site does
 % not have CM fonts, TeX will go haywire. Hence, we use \batchmode:
 %
  \immediate\write16{\sp@ce Loading in batchmode; please be patient.}
  \batchmode\input mtexsis\errorstopmode		% ..but in batchmode.
  \ATunlock\fi						% No ATlock yet
\else\immediate\write16{\sp@ce TeXsis.}\relax		% it is TeXsis
\fi

%           TeXsis commands may appear only below this line
%           ^^^^^^^^^^^^^^^^^^^^^^^^^^^^^^^^^^^^^^^^^^^^^^^

%%%%%%%%%%%%%%%%%%%%%%%%%%%%%%%%%%%%%%%%%%%%%%%%%%%%%%%%%%%%%%%%%%%%%%%%%%%%%%%
%%%%%%%%%%%%%%%%%%%%%%%%%%%%%%%%%%%%% Fonts %%%%%%%%%%%%%%%%%%%%%%%%%%%%%%%%%%%
%%%%%%%%%%%%%%%%%%%%%%%%%%%%%%%%%%%%%%%%%%%%%%%%%%%%%%%%%%%%%%%%%%%%%%%%%%%%%%%
%% Anil Trivedi, July 1993

\def\f@n{\fontname}\def\g@f{\global\font}
  \def\m@{   scaled \magstep}
\def\m@hf{   scaled \magstephalf}
%          ^ this space required

%% Does their Plain define \tenrm as amr10? If so, only seek AM fonts.
    \newif\ifoldfonts
    \edef\tenrmname{\fontname\tenrm}
    {\def\\#1{\catcode`#1=12 }\\a\\m\\r\\1\\0\gdef\AMRTEN{amr10}}  % need {}
    \ifx\tenrmname\AMRTEN\oldfontstrue\fi

%% The macro \F@NT accepts an order of preference, using the first font
%% that is available. It also provides for a "desperate option".
%
\def\everyF@NT{\relax}				% hook
\def\F@NT#1=#2#3#4#5#6{%1%
    \ifLocalProcessing				% I know what I have: just
    \g@f#1=#2%					% take the first choice!
    \else%
       \ifoldfonts%				% AM fonts:
       \batchmode\g@f#1=#5\errorstopmode%	% skip 1st/2nd/3rd choices
       \else%
          \ifForceTenPointFonts%		% explicit request
          \batchmode\g@f#1=#4\errorstopmode%	% skip 1st/2nd choices
          \else%
             \batchmode\g@f#1=#2\errorstopmode%	% Try 1st choice
             \ifx#1\nullfont%
             \batchmode\g@f#1=#3\errorstopmode%	% Try 2nd choice
             \ifx#1\nullfont%
             \batchmode\g@f#1=#4\errorstopmode%	% Try 3rd choice
             \fi\fi%
          \fi
          \ifx#1\nullfont%
             \batchmode\g@f#1=#5\errorstopmode%	% Try 4th choice
          \fi%
       \fi%
       \ifx#1\nullfont%
          \message{\sp@ce #1\def'ed to #6(desperate choice!)}
          \gdef#1{#6}%				% desperate choice
       \else%
          \message{\sp@ce #1set as \f@n#1}%
       \fi%
       \everyF@NT				% hook, used later too!
    \fi%
                      }%1%

\emsg{}\emsg{ \l@m Fonts\t@m}%
\ifLocalProcessing\emsg{\sp@ce(9, 10, 12, 14, 16 pt.)}
\else%
  \ifoldfonts
  \emsg{\sp@ce(your \NX\tenrm was amr10; I'll only look for old fonts)}\emsg{}%
  \else%
     \ifForceTenPointFonts%
     \emsg{\sp@ce(you asked for magnified 10pt fonts)}\emsg{}%
     \else%
     \emsg{\sp@ce(find the closest size, then scale if necessary)}%
     \fi%
  \fi
\fi%

%% 9 pt:
\def\ninefonts{%1%
  \F@NT\ninerm={cmr9}{}{cmr9}{amr9}{\tenrm}%
  \F@NT\ninei={cmmi9}{}{cmmi9}{ammi9}{\teni}%
  \F@NT\ninesy={cmsy9}{}{cmsy9}{amsy9}{\tensy}%
  \F@NT\nineex={cmex10}{}{cmex10}{amex10}{\tenex}%	% keep 10
  \F@NT\ninebf={cmbx9}{}{cmbx9}{ambx9}{\tenbf}%
  \F@NT\ninesl={cmsl9}{}{cmsl9}{amsl9}{\tensl}%
  \F@NT\ninett={cmtt9}{}{cmtt9}{amtt9}{\tentt}%
  \F@NT\nineit={cmti9}{}{cmti9}{amti9}{\tenit}%
  \skewchar\ninei='177\skewchar\ninesy='60\hyphenchar\ninett=-1%
  \gdef\ninefonts{\relax}%
              }%1%

%% 10 pt:                               	% most defined in Plain
\F@NT\tenss={cmss10}{}{cmss10}{amss10}{\tenrm}
\F@NT\tencsc={cmcsc10}{}{cmcsc10}{amcsc10}{\teni}
\let\tenp@int=\tenpoint
\def\tenpoint{\tenp@int\def\sc{\tencsc}\def\csc{\sc}}

%% 12 pt:
\def\twelvefonts{%1%
  \F@NT\twelverm={cmr12}{cmr10\m@1}{cmr10\m@1}{\f@n\tenrm\m@1}{\twelvess}
  \F@NT\twelvei={cmmi12}{cmmi10\m@1}{cmmi10\m@1}{\f@n\teni\m@1}{}
  \F@NT\twelvesy={cmsy10\m@1}{cmsy10\m@1}{cmsy10\m@1}{\f@n\tensy\m@1}{}
  \F@NT\twelveex={cmex10\m@1}{cmex10\m@1}{cmex10\m@1}{\f@n\tenex\m@1}{}
  \F@NT\twelvebf={cmbx12}{cmbx10\m@1}{cmbx10\m@1}{\f@n\tenbf\m@1}{\twelverm}
  \F@NT\twelveit={cmti12}{cmti10\m@1}{cmti10\m@1}{\f@n\tenit\m@1}{\twelvesl}
  \F@NT\twelvesl={cmsl12}{cmsl10\m@1}{cmsl10\m@1}{\f@n\tensl\m@1}{\twelveit}
  \F@NT\twelvett={cmtt12}{cmtt10\m@1}{cmtt10\m@1}{\f@n\tentt\m@1}{\twelverm}
  \F@NT\twelvess={cmss12}{cmss10\m@1}{cmss10\m@1}{amss10\m@1}{\twelverm}
  \F@NT\twelvecsc={cmcsc10\m@1}{}{cmcsc10\m@1}{amcsc10\m@1}{\twelveit}%
  \skewchar\twelvei='177\skewchar\twelvesy='60\hyphenchar\twelvett=-1%
  \gdef\twelvefonts{\relax}%
                }%1%
\let\twelvep@int=\twelvepoint
\def\twelvepoint{\twelvep@int\def\sc{\twelvecsc}\def\csc{\sc}}

%% 14 pt:
\def\fourteenfonts{%1% 				%last entry not completed yet
  \F@NT{\fourteenrm}={cmr12\m@1}{cmr10\m@2}{cmr10\m@2}{\f@n\tenrm\m@2}{}
  \F@NT{\fourteeni}={cmmi12\m@1}{cmmi10\m@2}{cmmi10\m@2}{\f@n\teni\m@2}{}
  \F@NT{\fourteensy}={cmsy10\m@2}{cmsy10\m@2}{cmsy10\m@2}{\f@n\tensy\m@2}{}
  \F@NT{\fourteenex}={cmex10\m@2}{cmex10\m@2}{cmex10\m@2}{\f@n\tenex\m@2}{}
  \F@NT{\fourteenbf}={cmbx12\m@1}{cmbx10\m@2}{cmbx10\m@2}{\f@n\tenbf\m@2}{}
  \F@NT{\fourteenit}={cmti12\m@1}{cmti10\m@2}{cmti10\m@2}{\f@n\tenit\m@2}{}
  \F@NT{\fourteensl}={cmsl12\m@1}{cmsl10\m@2}{cmsl10\m@2}{\f@n\tensl\m@2}{}
  \F@NT{\fourteenss}={cmss12\m@1}{cmss10\m@2}{cmss10\m@2}{amss10\m@2}{}
  \skewchar\fourteeni='177\skewchar\fourteensy='60%
  \gdef\fourteenfonts{\relax}%
                  }%1%

%% 16 pt:
\def\sixteenfonts{%1% 				%last entry not completed yet
  \F@NT{\sixteenrm}={cmr17}{cmr12\m@2}{cmr10\m@3}{\f@n\tenrm\m@3}{}
  \F@NT{\sixteeni}={cmmi12\m@2}{cmmi10\m@3}{cmmi10\m@3}{\f@n\teni\m@3}{}
  \F@NT{\sixteensy}={cmsy10\m@3}{cmsy10\m@3}{cmsy10\m@3}{\f@n\tensy\m@3}{}
  \F@NT{\sixteenex}={cmex10\m@3}{cmex10\m@3}{cmex10\m@3}{\f@n\tenex\m@3}{}
  \F@NT{\sixteenbf}={cmbx12\m@2}{cmbx10\m@3}{cmbx10\m@3}{\f@n\tenbf\m@3}{}
  \F@NT{\sixteenit}={cmti12\m@2}{cmti10\m@3}{cmti10\m@3}{\f@n\tenit\m@3}{}
  \F@NT{\sixteensl}={cmsl12\m@2}{cmsl10\m@3}{cmsl10\m@3}{\f@n\tensl\m@3}{}
  \F@NT{\sixteenss}={cmss17}{cmss12\m@2}{cmss10\m@3}{amss10\m@3}{}
  \F@NT{\sixteenssi}={cmssi17}{cmssi12\m@2}{cmssi10\m@3}{amssi10\m@3}{}
  \skewchar\sixteeni='177\skewchar\sixteensy='60%
  \gdef\sixteenfonts{\relax}%
                 }%1%
\sixteenpoint\twelvepoint   	% draw out all the font chatter right now
%%

%% Script Font "rsfs":
\emsg{}\emsg{ \l@m Looking for the script font "rsfs"\t@m}%
\newfam\Scrfam\newfam\scrfam				% Scr=12pt, scr=10pt
\batchmode\font\scrfont=rsfs10\errorstopmode		% see if it's there
\ifx\scrfont\nullfont					% rsfs10 not found...
   \emsg{\sp@ce Not found!!!}  \emsg{}
   \emsg{ \l@m You don't seem to have the script font "rsfs"; get it!}
   \emsg{ \l@m For now, the substitution \noexpand\scr -> \noexpand\cal
            will be made.}
   \emsg{ \l@m --Anil Trivedi} \emsg{}
   \def\s@r{\cal}					% define \s@r = \cal
\else							% rsfs10 exists
   \g@f\twelvescr=rsfs10\m@1 \g@f\eightscr=rsfs7\m@1 \g@f\sixscr=rsfs5\m@1
   \skewchar\twelvescr='177\skewchar\eightscr='177\skewchar\sixscr='177
   \textfont\Scrfam=\twelvescr\scriptfont\Scrfam=\eightscr
   \scriptscriptfont\Scrfam=\sixscr
   \g@f\tenscr=rsfs10 \g@f\sevenscr=rsfs7 \g@f\fivescr=rsfs5
   \skewchar\tenscr='177\skewchar\sevenscr='177\skewchar\fivescr='177
   \textfont\scrfam=\tenscr\scriptfont\scrfam=\sevenscr
   \scriptscriptfont\scrfam=\fivescr
   \def\scr{\fam\Scrfam}				% default is 12pt
   \def\s@r{\scr}					% define \s@r = \scr
   \message{\sp@ce Script font loaded.}
\fi
%%

%%%%%%%%%%%%%%%%%%%%%%%%%%%%%%%%%%%%%%%%%%%%%%%%%%%%%%%%%%%%%%%%%%%%%%%%%%%%%%%
%%%%%%%%%%%%%%%%%%%%%%%%%%%% Customizing TeXsis2.15 %%%%%%%%%%%%%%%%%%%%%%%%%%%
%%%%%%%%%%%%%%%%%%%%%%%%%%%%%%%%%%%%%%%%%%%%%%%%%%%%%%%%%%%%%%%%%%%%%%%%%%%%%%%

\emsg{}\emsg{ \l@m Loading Macros\t@m}

%%-----------------------------------------------------------------------------
%% Assorted stuff						   Anil Trivedi

\overfullrule=0pt\vbadness=10000\hbadness=10000\tolerance=400
\def\email#1{\smallskip\rm(#1)}

%% \parskip as a fraction of \baselineskip
\def\setp@rskip#1{%
   \parskip=#1\baselineskip
   \parskip=\the\parskip plus 1pt minus 1pt}
\def\setparskip{\setp@rskip{.2}}			% default=20%

%% Line spacing
% Multiply current line spacing
\long\def\multiplyspacing#1{%1%
   \baselineskip = #1\baselineskip
   \setRuledStrut\setTableskip				% TeXsis
   \setparskip}%1%					% sets \parskip
% Start with singlespaced before multiplying
\def\lsfactor#1{\singlespaced\multiplyspacing{#1}}
\def\normalspacing{\lsfactor{1.2}}			% 14x1.2 pt
			% Note 1.5

%% "List" environment parameters; change if needed
%\EnvLeftskip=2\parindent\EnvRightskip=\parindent
%\EnvTopskip=\medskipamount\EnvBottomskip=\medskipamount
\EnvTopskip=\smallskipamount\EnvBottomskip=\smallskipamount

%% In-text displayed quotes
\def\quotefont{\tenrm} \def\quotespacing{\normalspacing}
% stand-alone paragraphs:

% within a paragraph:
\def\quote#1{\description{#1}\quotefont\quotespacing}
\def\endquote{\enddescription\noindent}

%% Italic, Units
\def\It#1{{\it#1\/}}

%% Define \; for text mode too
%\def\,{\relax\ifmmode\mskip\the\thinmuskip\else\thinspace\fi}	% in TeXsis
\def\;{\relax\ifmmode\mskip\thickmuskip\else\,\,\fi}		% added

%% Journals
%% Note: \jnl'\HPA'13'269'1940' = \journal\HPA;13,269(1940)
\def\jnl{\bgroup\catcode`\.=\active\offparens\@jnl}\offparens
\def\@jnl'#1'#2'#3'#4'{%1%
   \ifEurostyle {#1} {\vol{#2}} (#4) #3\relax
   \else {#1} {\vol{#2}}, #3 (#4)\relax\fi\egroup
                      }%1%
%% References
\def\SetRefs{%1%				% Example output:
   \def\Ref##1{Ref.~\l@ref\use{Ref.##1}\r@ref}	% Ref. [7]    or Ref. 7
   \def\ref##1{\l@ref\use{Ref.##1}\r@ref}	% [7]         or 7
   \def\xref##1{\use{Ref.##1}}			% just 7
   \def\citerange{\refrange}			% [7-9]       or $^{7-9}$
   \def\Refrange##1##2{%			% Refs. [7-9] or Refs. 7-9
      Refs.~\l@ref\use{Ref.##1}--\use{Ref.##2}\r@ref}
   \ifsuperrefs					% superscripts
      \def\l@ref{}\def\r@ref{}
   \else					% brackets
      \def\l@ref{[}\def\r@ref{]}
  \fi}%1%					% end \SetRefs
\let\sup@rr@fstrue=\superrefstrue
\let\sup@rr@fsfalse=\superrefsfalse
\def\superrefstrue{\sup@rr@fstrue\SetRefs}
\def\superrefsfalse{\sup@rr@fsfalse\SetRefs}
\superrefsfalse					% Important default

%% Equation numbers
%
\def\Eq#1{Eq.~($\use{Eq.#1}$)}			% Eq. (7)
		% Eqs. (7)
\def\Ep#1{($\use{Eq.#1}$)}          		% (7)
 			% (7)
 	% 7
%
% The following replaces \advance by
% \global\advance (bug-fix)
%
\def\@use#1{\endgroup\stripblanks @#1@\endlist
   \XA\ifx\csname\tok\endcsname\relax\relax
     \emsg{\l@m UNDEFINED TAG #1 ON PAGE \folio.}%
     \global\advance\@BadTags by 1
     \@errmark{UNDEF}\edef\tok{{\bf\tok}}%
   \else\edef\tok{\csname\tok\endcsname}%
   \fi\tok}%
\def\make@refmark#1{%1%
   \testtag{Ref.#1}\ifundefined
     \emsg{\l@m UNDEFINED REFERENCE #1 ON PAGE \number\pageno.}%
     \global\advance\@BadRefs by 1
     \xdef\@refmark{{\tenbf #1}}\@errmark{REF?}
   \else\xdef\@refmark{\csname\tok\endcsname}%
   \fi}%1%

%%----------------------------------------------------------------------------
%% Informal Stuff						  Anil Trivedi

% Stuff between \Informal and \endInformal is ignored if \formal or
% \noInformal has been declared. \Manuscript and \PhysRev are formal.

\def\keepInformal{\def\Informal{\relax}\def\endInformal{\relax}}
\def\noInformal{\long\def\Informal##1\endInformal{\relax}
                \def\endInformal{\relax}}

\keepInformal							% default

% Informal abstract
\long\def\informalabstract#1\endinformalabstract{%1%
   \Informal\informalabstractskip			% def'd later
   \noindent(\It{Informal Abstract.}\ #1)
   \endInformal\def\endinformalabstract{\relax} }%1%

% Quote(s) at the bottom of last page (informal)
% Use \pj...\endpj pairs; \pjauthor is optional
\newif\iffirstpj\firstpjtrue
\long\def\pj#1\endpj{%1%
   \Informal
   \iffirstpj\vskip0pt plus1filll\firstpjfalse
   \else\vskip1.5\baselineskip\fi
   \bgroup \def\\{\par}					% \\ is linebreak
   \def\pjauthor{\twelvess\medskip---\ }
   \def\endpj{\relax}
   \twelvepoint\parskip=0pt\sl
   \flushright #1 \endflushenv
   \egroup\endInformal}%1%

%%-----------------------------------------------------------------------------
%% Titlepage.		 					   Anil Trivedi

%% The very top; say \verytop{..whatever..}
\def\verytop#1{%1%					%% \top is taken
    \gdef\VERYT@P{%2%					%% called later
      \line{\hfil\vbox to 0pt{\vss\hbox{\twelvess#1}}\hfil}
      \vskip12pt }%2%
    \gdef\verytop##1{}	}%1%				%% used only once!
\def\VERYT@P{\vglue-\topskip}				%% need as default
\ifLocalProcessing
  %1%			%% in ~/Tex/ps
\else
  
\fi

%% Number assigned by HEP-LAT or HEP-TH, etc.
\def\eprintcode#1{\gdef\@EPRINTcode{#1}}\eprintcode{}	%% default=empty

%% Manuscript date, called by \Manuscript
\def\manuscriptdate#1{\gdef\@MANUSCRIPTdate{#1}}
\manuscriptdate{\today}					%% default=autodate
% (Better to use manually, so a record remains)
%
%% Journal Name, called by \Manuscript
\def\journalname#1{\gdef\@JOURNALname{#1}}		%% \journal is taken
\journalname{}						%% default=empty
%
%% Manuscript number of a journal, called by \Manuscript
\def\manuscriptcode#1{\gdef\@MANUSCRIPTcode{#1}}
\manuscriptcode{}					%% default=empty

%%-----------------------
%% Abstract-length is the big unknown: distribute glue between the objects
%% making up the coverpage; if necessary try to keep everything on one page
%% by letting the baselineskip in the abstract body shrink to singlespaced.

\newdimen\abstractshr@nk				% used in \abstract
\def\MyTitleP@ge{%1%
%   \emsg{}\emsg{ \l@m Starting title page.}\emsg{}
   \def\banner{%2%
      \VERYT@P
      \hrule height0.2pt\vskip1.5pt\hrule height0.2pt	% changed -AKT
      \vskip4pt\relax
      \line{\twelvepoint\rm\@LEFT \hfil \@RIGHT}
      \vskip-\baselineskip\centerline{\@CENTER}		% really center it
      \vskip3pt%
      \hrule height0.2pt\vskip1.5pt\hrule height0.2pt%
              }%2%
   \def\title{%3%
      \endmode\vskip 1cm plus 1fil			% 1cm+ 1fil glue
      \mark{Title Page\NX\else Title Page}
      \bgroup\setparskip				% 20%; defined above
      \let\endmode=\endtitle
      \def\\{\par}					% added -AKT
      \center\Tbf}%3%
   \let\@author=\author
   \def\author{\vskip0pt plus 0.1fil \@author}		% 0.1fil glue
   \def\abstract{%4%
      \endmode\bigskip\bigskip
      \vskip0pt plus .5fil				% 0.5fil glue
      \centerline{A\,B\,S\,T\,R\,A\,C\,T}		% \, added -AKT
      \medskip\bgroup
      \let\endmode=\endabstract
      \narrower\narrower\normalspacing			% normalspacing...
      \abstractshr@nk=\baselineskip			% ...but can shrink
      \advance\abstractshr@nk by -\normalbaselineskip	% to singlespaced
      \baselineskip=\the\baselineskip minus \abstractshr@nk
      \noindent}%4%
   \def\informalabstractskip{\vskip 0.75\baselineskip}
   \def\pacs##1{%5%
      \vskip\baselineskip\vskip0pt plus 2fil		% 2fil glue
      \centerline{PACS numbers: ##1}
      \smallskip}%5%
   \def\endtitlepage{%6%
      \vskip0pt plus 1fil				% +usual 1in
      \preprintonly{\vskip 0pt plus 0.5fil
                    \leftline{{\tenpoint\TeXsis~\fmtversion}}}
      \endmode\eject\egroup
%      \emsg{}\emsg{ \l@m Starting main paper.}\emsg{}
                    }%6%
   \def\MyTitleP@ge{}					% used only once
                       }%1%				% end \MyTitleP@ge

%%-----------------------------------------------------------------------------
%% Styles.		 					   Anil Trivedi

%% For PRL-style sections in preprint; changed in \PhysRev
\def\runningsectionskip{\bigskip}

%% All style definitions eliminated as soon as one is used
\def\nomor@styl@s{%1%
      \gdef\@preprint{}\gdef\preprint{}
      \gdef\@PhyRev{}\gdef\PhyRev{}
      \gdef\@paper{}\gdef\paper{}\gdef\Manuscript{}
                 }%1%

%% Preprint  (separate coverpage)
\newif\ifpreprint\let\@preprint=\preprint
\def\preprint{%1%
      \preprinttrue
      \def\@LEFT{\@PUBdate}
      \def\@CENTER{\@EPRINTcode}
      \def\@RIGHT{\@DOCcode}
      \@preprint\normalspacing				% 16.8 pt, parskip set
      \def\Tbf{\sixteenpoint\bf}
      \MyTitleP@ge					% AFTER \@preprint
      \def\pacs##1{\vskip0pt plus 3fil}			% disable; some stretch
      \nomor@styl@s					% only one style used
             }%1%					% End \prerint

%% Manuscript
%% Consider: \raggedright, \raggedbottom??
\newif\ifManuscript
\def\Manuscript{%1%
      \Manuscripttrue
      \def\@LEFT{\@MANUSCRIPTdate}
      \def\@CENTER{\@JOURNALname}
      \def\@RIGHT{Manuscript\,\#~\@MANUSCRIPTcode}
      \def\normalspacing{\lsfactor{1.75}\parskip=0pt} % 24.5 pt
      \@preprint\normalspacing
      \def\Tbf{\sixteenpoint\bf}
      \MyTitleP@ge					% AFTER above
%      \raggedright\raggedbottom			% Don't like it!
      \noInformal\FiguresLast\nomor@styl@s
               }%1%					% End \Manuscript

%% Paper (no coverpage)
\newif\ifpaper\let\@paper=\paper
\def\paper{%1%
      \papertrue\@paper\normalspacing
      \def\Tbf{\sixteenpoint\bf}
      \def\pacs##1{\relax}				% disable
      \let\@abstract=\abstract
      \def\abstract{\@abstract\noindent}
      \def\informalabstractskip{\medskip}
      \def\endtitlepage{\endmode\goodbreak
                        \vskip .5in\egroup}
      \nomor@styl@s					% only one style used
          }%1%						% End \paper

%% PR style output (needs full TeXsis)
\newif\ifPhysRev\let\@PhysRev=\PhysRev\def\PRcomma{}	% empty in other styles
\def\PhysRev{%1%
      \PhysRevtrue\@PhysRev\noInformal
      \def\scr{\fam\scrfam}			% smaller size
       \let\@title=\title
      \def\title{\@title\unobeylines		% suspend \obeylines, but
                 \def\\{\par}}			% \\ will force a linebreak
      \def\PRcomma{,\space}			% , here; or use \\
      \def\pacs##1{%2%
         {\leftskip=1in\noindent
          \ninepoint PACS numbers: ##1		% PR style
          \smallskip} }%2%
      \EnvLeftskip=20pt\EnvRightskip=10pt	% indents in lists
      \def\quotefont{\ninerm}
      \def\quotespacing{\singlespaced}
      \SetRefs					% defined above
      \def\vol##1{{\bf ##1}}			% PR style
      \def\refFormat{\ninepoint}		% hook in \Listreferences
      \def\@refitem##1##2{\message{##1.}%%	% Spaces touchy
         \refskip\noindent\hskip-\refindent
         \hbox to\refindent{\hss[##1]\SP{.3em}}%% [7] in Reflist
                      ##2}%
      \def\runningsectionskip{\smallskip}	% "sections" in PRL
      \nomor@styl@s				% only one style declaration
            }%1%				% End \PhysRev
% Note: \@refitem determines how ref. #s are printed in the final list.

% Material to be included only in a particular style
\long\def\preprintonly#1{\ifpreprint#1\fi}
\long\def\Manuscriptonly#1{\ifManuscript#1\fi}

%
% Material to be excluded from a particular style

\long\def\Manuscriptexclude#1{\ifManuscript\relax\else#1\fi}

%%-----------------------------------------------------------------------------
%% Endgame							   Anil Trivedi

%% Final listing of references: Eliminate separate reporting on unresolved
%  \cite's (see next macro); no extra space between refs; use [7] instead
%  of 7., if text has [] citations.

\def\ListReferences{%1%				% For default see TXSrefs.tex
   \errorstopmode				% In case we are in batchmode
   \emsg{}\emsg{ \l@m Final reference list\t@m}
   \p@nctwrite.\relax
   \emsg{}
   \@refwrite{\@comment>>> EOF \jobname.ref <<<}
   \immediate\closeout\reflistout
   \begingroup\catcode`@=11\offparens\unobeylines
   \def\refskip{\vskip0pt plus 1pt}		% no extra space, 1pt stretch
   \ifsuperrefs						% superscript citations
       \setbox\tempbox\hbox{\the\refnum.\quad}		% set indent
       \def\@refitem##1##2{\message{##1.}%%%		% 7. on the screen
            \refskip\noindent\hskip-\refindent
            \hbox to \refindent {\hss ##1.\quad}%%%	% 7. in the list
            ##2}%
   \else						% [] citations
       \setbox\tempbox\hbox{[\the\refnum]\quad}		% set indent
       \def\@refitem##1##2{\message{##1.}%%%		% 7. on the escreen
            \refskip\noindent\hskip-\refindent
            \hbox to \refindent {\hss [##1]\quad}%%%	% [7] in the list
            ##2}%
   \fi
   \refindent=\wd\tempbox \leftskip=\refindent
   \parindent=\z@ \def\reference{\@noendref}
   \refFormat						% hook for changes
   \Input\jobname.ref\relax\vskip0pt\endgroup\emsg{}
                   }%1%				% End ListReferences

%% Combined checking of unresolved tags AND \cite's
%
\def\checktags{%1%
   \newlinechar=`\^^J
   \advance \@BadTags by \@BadRefs		% only bother with the sum
   \ifnum\@BadTags=\z@
   \immediate\write16{%2%
  ^^J
  \l@m All labels and references have been successfully processed.^^J
  \l@m Please print the DVI file \jobname.dvi; afterwards, you may wish^^J
  \l@m to delete the auxiliary files \jobname.aux, \jobname.ref,
       and \jobname.log.^^J
  \l@m (In case of printing problems, see opening comments in \jobname.tex.)
  ^^J
                  }%2%
   \else
   \immediate\write16{%3%
  ^^J
  \l@m Attention! Don't print anything. Don't delete any file.^^J
  \l@m There were \the\@BadTags\space unresolved labels:
       you need a second run.^^J
  \l@m Please process \jobname.tex once again!
  ^^J
                  }%3%
   \fi        }%1%				% End checktags

%%%%%%%%%%%%%%%%%%%%%%%%%%%%%%%%%%%%%%%%%%%%%%%%%%%%%%%%%%%%%%%%%%%%%%%%%%%%%%%
%%%%%%%%%%%%%%%%%%%%%%%%%%%%%%%%%%% Aliases %%%%%%%%%%%%%%%%%%%%%%%%%%%%%%%%%%%
%%%%%%%%%%%%%%%%%%%%%%%%%%%%%%%%%%%%%%%%%%%%%%%%%%%%%%%%%%%%%%%%%%%%%%%%%%%%%%%

%% Since these are always loaded, don't appropriate any Plain TeX names.
%% Do that "locally", in individual files.

%%-----------------------------------------------------------------------------
%% Journals							   Anil Trivedi

\def\AP{Ann.\ Phys.\ (N.Y.)}		
\def\CMP{Commun.\ Math.\ Phys.}

\def\NP{Nucl.\ Phys.}			\def\PL{Phys.\ Lett.}

\def\PRD{Phys.\ Rev.\ D}

%%-----------------------------------------------------------------------------
%% Greeks							   Anil Trivedi

\def\al{\alpha}

	 \def\Ga{{\mit\Gamma}}

\def\th{\theta}

\def\la{\lambda} 	
\def\rh{\rho}

\def\ph{\phi}	 		  

\def\ps{\psi}

%%-----------------------------------------------------------------------------
%% Uppercase bolds; cals; scripts.				   Anil Trivedi

\def\sA{{\s@r A}}  \def\sB{{\s@r B}}  \def\sC{{\s@r C}}  \def\sD{{\s@r D}}
\def\sE{{\s@r E}}  \def\sF{{\s@r F}}  \def\sG{{\s@r G}}  \def\sH{{\s@r H}}
\def\sI{{\s@r I}}  \def\sJ{{\s@r J}}  \def\sK{{\s@r K}}  \def\sL{{\s@r L}}
\def\sM{{\s@r M}}  \def\sN{{\s@r N}}  \def\sO{{\s@r O}}  \def\sP{{\s@r P}}
\def\sQ{{\s@r Q}}  \def\sR{{\s@r R}}  \def\sS{{\s@r S}}  \def\sT{{\s@r T}}
\def\sU{{\s@r U}}  \def\sV{{\s@r V}}  \def\sW{{\s@r W}}  \def\sX{{\s@r X}}
\def\sY{{\s@r Y}}  \def\sZ{{\s@r Z}}

%%-----------------------------------------------------------------------------

\def\apriori{{\it a~priori\/\ }}

\def\circa{{\it circa\/\ }}

\def\SP#1{\hskip #1\relax}			% \quad=\SP{1em}
\def\:{\mskip -1.5mu}				% 1/2 of \!
% Notes: \? is taken in TeXsis; \mskip must be be in math-units (mu=em/18)

\def\D{\sD}					% functional measure
				% differential; don't use \d
\def\pl{\partial}				% Phys. Lett. is \PL
\def\dd#1{{{\pl\,}\over{\pl#1}}}		% \dd{t} is time-derivative
 				% use: {A_1}\dg
\def\R{[-\infty, \infty\,]}			% use \Re for Gothic R
\def\BZ{[-\pi\:/\:\ell, \pi\:/\:\ell\,]}	% Brillouin Zone

\def\intx{\int \!{\rm d}^4\:x\,}

\emsg{\sp@ce Loaded.}\emsg{}

\ATlock						% IMPORTANT, VERY IMPORTANT!
%%

%%%%%%%%%%%%%%%%%%%%%%%%%%%%%%%%%%%%%%%%%%%%%%%%%%%%%%%%%%%%%%%%%%%%%%%%%%%%%%%
%%%%%%%%%%%%%%%%%%%%%%%%%%%%%%%%% File l3.tex %%%%%%%%%%%%%%%%%%%%%%%%%%%%%%%%%
%%%%%%%%%%%%%%%%%%%%%%%%%%%%%%%%%%%%%%%%%%%%%%%%%%%%%%%%%%%%%%%%%%%%%%%%%%%%%%%
%% Anil Trivedi, October 1993

\texsis\preprint
\pubdate{October 1993}
\eprintcode{hep-lat/9309012}
\pubcode{AKT-93-L3}

%%-----------------------------------------------------------------------------

\titlepage

\title							% \obeylines
The Nielsen-Ninomiya ``No-Go'' Theorem
Is False
\endtitle

\author
Anil K. Trivedi
The Enrico Fermi Institute, The University of Chicago\PRcomma
5640 South Ellis Avenue, Chicago, Illinois 60637, USA
\email{E-mail: trivedi@yukawa.uchicago.edu}
\endauthor

\abstract
The Nielsen-Ninomiya no-go theorem asserts that chiral Weyl~(``neutrino'')
fields cannot exist on lattices. However, the actual mathematical arguments
advanced in the theorem fail to make that case. The theorem leaves the
problem of lattice neutrinos completely open; it identifies no obstacle
which would prevent us from discretizing chiral theories.%
\informalabstract
Have you too, like every second person in the field, been citing the
Nielsen-Ninomiya theorem but have not read or analyzed it? Then, this paper
will wake you up. Reading it is a prerequisite to doing any good work in
chiral lattice physics.\,%
\endinformalabstract
\endabstract


\endtitlepage

%%-----------------------------------------------------------------------------

\def\newsec#1{\goodbreak\runningsectionskip\noindent{\bf#1}}
\def\NN{Nielsen-Ninomiya\ }
\def\t{\hat t}\def\l{\ell}
\def\xvec{{\vec x}}
\def\lowl{{\lower2pt\hbox{${\scriptstyle \l}$}}}
\def\lowI{{\lower2pt\hbox{${\scriptstyle I}$}}}
\def\intsum{\int \!{\rm d} t \sum_{\xvec\,} \l^{3\,}}
\def\ahat{{\hat a}}	\def\bhat{{\hat b}}	\def\hhat{{\hat h}}
\def\ghat{{\hat g}}	\def\lhat{{\hat\la}}	\def\rhat{{\hat\rh}}
\def\Hhat{{\hat H}}	\def\Fhat{{\hat F}}	\def\Khat{{\hat K}}
\def\psbar{\bar\ps}
\def\Dpsi{\D\:\ps\D\:\psbar}
\def\intpsi{\int\!\Dpsi}
\def\Repn#1{\Ga \bigl\{ #1 \bigr\} }		% displayed eqns
\def\repn#1{\Ga \bigl\{ #1 \bigr\} }		% in-text eqns; same for now

%%-----------------------------------------------------------------------------
%
\newsec{1.~Introduction.}
The \NN ``no-go'' theorem, put forth in two papers titled
``Absence\ Of\ Neutrinos\ On\ A\ Lattice''%
\reference{NN1}
H.~B.~Nielsen and M.~Ninomiya, \jnl'\NP'B185'20'1981'
% ; erratum in \jnl'\NP'B195(E)'54'1982'
\endreference
\reference{NN2}
H.~B.~Nielsen and M.~Ninomiya, \jnl'\NP'B193'173'1981'
\endreference\space
(Friedan\reference{friedan}D.~Friedan, \jnl'\CMP'B85'481'1982'\endreference
gave a third proof) asserts that chiral Weyl~(``neutrino'') fields cannot
exist on lattices:
if so, we could not discretize chiral theories such as the Standard Model;
the ensuing conceptual issues have been compared to those Einstein faced
with relativity%
\reference{NN3}
H.~B.~Nielsen and M.~Ninomiya, Niels Bohr Institute preprint NBI-HE-91-08
\endreference.\space
\Ignore
It comes to us with enviable credentials%
\reference{credentials}
For example, each of the cited proofs invokes a different branch of
mathematics: \Ref{NN1} employs homotopy theory, \Ref{NN2} differential
topology, and \Ref{friedan} differential geometry
\endreference\endIgnore
It has never received the critical scrutiny commensurate with
the importance of its conclusions. % or influence.

My purpose here is to show that the \NN {no-go} theorem is false%
\reference{clarification}
This reassessment is a painful need of the hour. The \NN theorem has emerged
as a ``choke~point'' in physics: It casts doubt on the existence of chiral
theories. It would obstruct the numerical study of elementary-particle models,
even if adequate computing machines and algorithms became available. It would
obstruct any serious understanding of a fundamental length and by implication
possibly of quantum gravity as well (see \Ref{NN3}). You cannot refute it with
counterexamples or countertheorems: absent an understanding of what is wrong
where, a no-go theorem plus counterexamples or countertheorems only add up to
an unresolved paradox. The evolution of the subject, the intellectual force of
the theorem, and its uncritical acceptance have created a powerful paradigm
that must be faced in its own right
\endreference
\reference {technical}
As in \Refrange{NN1}{friedan}, time is left continuous.
Take the spatial lattice to be three-dimensional,
with the spacing $\l$ and the unit vectors $\vec e_j$.
Defining
$\t_j^N f(\xvec) \!\equiv\! f(\xvec \!+\! N\l\vec e_j)$
for integer $N$, a lattice operator can be expanded as
$\lhat \!=\! \sum_{\vec{n}} \,\la_{\vec{n}} \>
  \t_1^{\,n_1} \, \t_2^{\,n_2} \, \t_3^{\,n_3}$;
it is local if the expansion terminates.
Define its conjugate
$\lhat^* \!=\! \sum_{\vec{n}} \,\la_{-\vec{n}}^*\>
  \t_1^{\,n_1} \, \t_2^{\,n_2} \, \t_3^{\,n_3}$,
and the \It{left-}action of operators
$f(\xvec) \, \t_j^N \!\equiv\! f( \xvec \!-\! N \l {\vec e}_j )$.
This yields two identities:
$\overline {\lhat \ps} \!\equiv\! \psbar \lhat^*$,
and
$\sum_{\vec x}\,\psbar (\lhat\ps) \!=\! \sum_\xvec\,(\psbar\lhat)\ps$;
the latter is the lattice analog of integration by parts
\endreference:\space
While its mathematical content produces a useful insight, its principal
conclusion as well as its assertions concerning the Standard Model are
incorrect.
% the lattice formulation of chiral quantum field theories is an open problem.
Its arguments are irrelevant to the existence or the non-existence of
lattice neutrinos; the insight we can salvage from them is a weak constraint
on parameterization of such theories.
% Though useful, the constraint is weak: of the uncountably infinite number
% of nonlocal representations that every theory possesses, one is identified.
% Despite their impressive mathematical span and depth,
% neither the \NN proofs\cite{NN1}\cite{NN2} nor Friedan's proof\cite{friedan}
No proof given to date\citerange{NN1}{friedan} contains any more information
on lattice neutrinos.

The following analogy will help. Imagine a ``gedanken-society'' which
is used to linear equations, polar coordinates, and the circles $r\:=\:c$.
Time comes when cartesian coordinates would be convenient. However, the
society in question is unable to find the new equation of a circle and
summarizes that experience in a ``non-existence'' theorem titled
``Absence of circles in cartesian coordinates'': {``We prove a no-go theorem
for circles in cartesian coordinates under very general and mild assumptions.
Since nonlinear equations carry the risk of bad geometric properties
a circle does not have, we can avoid them and examine the general curve
$ax\:+\:by\:=\:c$ where $a,b,c\,$ can be anything. Here are several proofs
that whichever curve you pick, its net curvature is necessarily zero.
Hence circles cannot exist in cartesian coordinates.''}
This gedanken-theorem misformulates the problem and misinterprets whatever
result remained. It gets there by losing sight of geometric objects~(circles)
or properties~(curvature) and focusing on parameterized ones~(equations,
linearity): instead of asking if some parameterization describes circles,
one should ask what kind can. And by ``equivalencing'' all nonlinear
equations: most are unsuitable, but \It{one} of them is just what is needed.
It uncovers a fact: unlike the linear polar equation $r\:=\:c$, the linear
cartesian equation $ax\:+\:by\:=\:c$ does not describe circles. That is
interesting, but a far cry from the non-existence of circles or curvature
in cartesian coordinates. As it draws and promotes an incorrect conclusion,
the theorem is false---though perhaps more informative than many true~ones.

The \NN no-go theorem is false and informative in the same sense. At first
you may find my analogy unfair---polar and cartesian coordinates obviously
span the same plane, whereas a continuum and a lattice are different
spaces---but it is a valid one. Quantum field theories possess momentum
representations and the continuum momenta $\in\!\R$ and the lattice momenta
$\in\!\BZ$ are as much in one-to-one correspondence as are polar and
cartesian coordinates. This, ``Continuum Lattice Duality''%
\reference{tobe}
A.~K.~Trivedi, to be published
\endreference,\space
tells us that if a theory exists in one space, it exists in the other too,
however different the look. It is our first hint that the no-go theorem
must have erred somewhere.

This article explains where. The discussion here, as in the \NN theorem,
is independent of specific formalisms. I shall construct one class of
lattice theories of neutrinos in a companion article%
\reference{l4}%
\Manuscriptexclude{A.~K.~Trivedi, hep-lat/9309013}%
\Manuscriptonly{Physical Review Letters manuscript LJ5262 \It{(the Editor
is requested kindly to add the reference)\rm}}
\endreference.\space

%%-----------------------------------------------------------------------------
%
\newsec{2.~The Theorem.}
The \NN theorem starts with a spatial lattice. Deeming the actions
$$
%\SP{3em}
S_{\lhat}[\psbar,\ps]=
\intsum \psbar [i \lhat \dd{t} - {\hat G}] \ps
%\SP{1.5em} \lparen \lhat  \not= {\rm constant} \rparen
\EQN S.g $$
as likely to give rise to unphysical spectra, it restricts the search
for physically relevant theories to local translationally-invariant
real actions of the form
$$
S_1[\psbar,\ps]=
\intsum\psbar[i \dd{t}-\Hhat] \ps
\EQN S.NN $$
possessing the correct continuum limit%
\reference{TI+CL}
{}From now on the translational invariance and the correct continuum limit
will be assumed of all actions
\endreference.\space
It finds that such a formalism cannot describe a system of non-zero chirality:
if an internal symmetry generates a conserved quantized charge whose density
is local and quadratic in $\ps$,
% (e.g., the lepton number or the weak hypercharge), %%%
then its irreducible representations possess zero net
chirality\refrange{NN1}{friedan}. It concludes that
\quote{\SP{2.5em}} %% argument determines indent
\itm{(A1)} `` ... the weak interaction cannot be put on the lattice''%
\reference{quote1}
\Ref{NN1}, p.~21
\endreference,\space
\endquote
which was its \It{raison d'\,\^etre}; a secondary conclusion is that
chirally invariant strong interaction models are also not possible.

For another hint that (A1) is not correct, see where it leads:
since weak interactions have been observed, the spacetime \It{does~not}
possess a lattice structure at {any} scale. However, since such a
structure at an untested scale is an open experimental question,
that conflicts with ``the~Scientific Method''.
In fact, the authors are aware of this conflict,
\quote{\SP{2.5em}}
\itm{(A2)} ``... if nature were indeed built on a lattice it should be
                possible to realize Weyl fermions ...
              % with different quantum numbers for left- and right-handed ones
                this is, however, just what cannot be done''%
\reference{quote2}
\Ref{NN1}, p.~23
\endreference,\space
\endquote
though perhaps insufficiently disturbed by it; if sound, conclusions such as
(A1) would invalidate the Scientific Method.

%%-----------------------------------------------------------------------------
%
\newsec{3.~A Theory and Its Representations.}
A physical theory predicts certain observables;
a representation of that theory is a specific parameterization of
calculational details.
% Theories come in many representations since intermediate steps are
% unimportant and can be changed if compensating changes are made elsewhere.%%
A quantum field theory, its observables obtainable from integrals
$$
F(u') = \int\!\sD\:\ph \; f\:[\ph\:(u')]\;
        \exp {i\!\int\!{\rm d}u\;\sL[\ph\:(u)]},
\EQN repn $$
admits arbitrary parameterization at two levels: both $u$ and $\ph$
are {dummy} variables.
%
% The former parameterizes the inner
% (ordinary) integration and determines the ``inner'' representation of
% the theory; the latter parameterizes the outer (path) integration and
% determines the ``outer'' or ``functional'' representation.
%
A continuum theory consists of instructions like
\def\qA#1{ q_0^A \lparen\,#1\rparen }
\def\Exp#1{\exp i\!\intsum\psbar#1\ps}
\def\path#1#2#3{{
           {\raise1pt\hbox{$\intpsi \; \qA{\overline{#1\ps},#2\ps}\;\Exp{#3}$}}
		\over
           {\lower1pt\hbox{$\intpsi \, \Exp{#3} $} }
               }}
$$
Q_0^A =  {{\intpsi \; \qA{\psbar,\ps} \>
%e^{i\!\intx\psbar\Khat_0\ps}}\over{\int\Dpsi\,e^{i\!\intx\psbar\Khat_0\ps}
\exp i\!\intx\psbar\Khat_0\ps}\over{\int\Dpsi\,\exp i\!\intx\psbar\Khat_0\ps
}}\, .
\EQN Q.cont $$
Let $\repn{\ahat_\lowl,\bhat_\lowl;\,\Khat_\lowl}$
denote its general discretization:
$$
Q_\lowl^A = \path{\ahat_\lowl}{\bhat_\lowl}{\Khat_\lowl} \, .
\EQN Q.lat $$

Now, \Eq{S.NN} only defines a parameterization-dependent object in terms of
dummy fields. A complete prescription, when finally given, would look like
$$
Q_\lowl^A = \path{\hhat_1}{\hhat_2}{\lparen i\dd{t}\!-\!\Hhat \rparen} .
\EQN repn.h $$
This would be a theory, {in the representation\/
$\repn{\hhat_1,\hhat_2;\, i\dd{t}\!-\!\Hhat}$}
which I shall call ``hamiltonian''.
However, just as a calendar-scheme should not be confused with history
or a coordinate-system with the space, you should not confuse this (or
any other) representation with the theory itself; $\psbar$ and $\ps$ are
dummy and may be changed.

For example, let $\sT\:(\rhat_1,\rhat_2)$ stand for the change % of variables
$(\psbar,\ps)\:\rightarrow\:(\psbar',\ps')$ where
${\psbar}\:=\:\overline{\rhat_1\ps'}\:=\:\psbar' \rhat_1^{\,*}$ and
$\ps\:=\:\rhat_2\ps'$
(assuming $\rhat_a$ to be linear and field-independent, the Jacobian
$\sD\:\ps\sD\:\psbar/\sD\:\ps'\:\sD\:\psbar'$ is a constant).
Applying it to \Eq{repn.h} yields (the primes are omitted):
$$
Q_\lowl^A = \path{\ghat_1}{\ghat_2}{\Fhat} ,
\EQN repn.g $$
where $ \ghat_a \!=\! \hhat_a \rhat_a$,
and $\Fhat \!=\! \rhat_1^{\,*} (i\dd{t} \!-\! \Hhat) \rhat_2$.
In general, representations transform as
$$
\sT\:(\rhat_1, \rhat_2 ) \, {\scriptstyle\circ} \;
\Repn{\ahat,\bhat;\,\Khat} =
\Repn{\ahat \rhat_1,\bhat \rhat_2;\,\rhat_1^{\,*} \Khat \rhat_2} ,
\EQN rep.transform $$
%
%this transformation preserves locality if (but not only if) $\rhat_1$
%and $\rhat_2$ are local.
%
which also means that the formalisms $\Repn{\ahat,\bhat;\,\Khat}$ and
$\Repn{\ahat \rhat_1,\bhat \rhat_2;\,\rhat_1^{\,*} \Khat \rhat_2}$
represent the same theory.

The physics, being about observables and not dummy variables,
is contained neither in the action $S[\psbar,\ps]$,
nor in the path-transcription $q^A(\psbar,\ps)$ of the observables,
but in their interplay%
\reference{misunderstanding}
Even in continuum, the action alone does \It{not} determine
a theory; e.g., recall the initial confusion that surrounded the
Schr\"odinger equation until Born's ``interpretation''---a prescription
for computing observables from the Schr\"odinger field---came along
\Ignore
  However, nowadays an \apriori convention for obtaining
  observables from fields is always assumed; having evolved from experience
  with continuum theories it \It{is} suitable there, but in a new arena like
  the lattice, an \apriori commitment to this or any other convention,
  without understanding the pros and cons of various choices, would be unwise
\endIgnore
\endreference:\space
\It {neither has any physical relevance without the other.}
Various representations of the same theory can differ in their actions,
in path-transcription of the observables, and in other attributes.
Clearly, the \It{manifest} hamiltonian structure and
the \It{manifest} locality are representation-dependent features%
\reference{representationology}
All theories possess a hamiltonian representation.
All possess nonlocal representations.
A theory need not possess local representations (most do not),
and even if it does, % (then the \It{theory} is local)
most of its \It{representations} are still nonlocal
\endreference.

%%-----------------------------------------------------------------------------
%
\newsec{4.~Critique of the Theorem.}
The summary of my critique of the no-go theorem is as follows:
The theorem misformulates the problem.
Its arguments contain a fatal technical defect.
It misinterprets whatever result remained.

For a query into the existence of lattice neutrinos, the theorem's scope
is unpromising, indeed unacceptably narrow: it avoids almost all local
lattice actions  there are [the form~\Ep{S.g}], and examines one lone
 form~\Ep{S.NN}.
There are $2^{\aleph_0}$ operators $\lhat$, of which local ones fall
into $\aleph_0$ distinct forms%
\reference{aleph}
Here $\aleph_0$ is the number of all integers, $2^{\aleph_0}$ that of
all real numbers
\endreference;\space
of those the authors exclude all but one.
% The theorem's assumptions are not ``mild'';
% it is difficult to imagine a stronger restriction.

While one can hardly better the authors's own demonstration of how
unsuitable their chosen form~\Ep{S.NN} was for describing neutrinos,
you can also view this restriction in the light of the following.
[i]~In the continuum, it is an \It{experimental} fact that we can describe
physical systems by local hamiltonian actions. With no such information
on the lattice, it would be prudent to work with the most general form
admissible.
[ii]~A non-existence proof has to be especially comprehensive;
e.g., a proof that some equation did not describe circles or another
one neutrinos would not prove the non-existence of anything.
[iii]~The authors offer very little justification for excluding the general
form~\Ep{S.g}---and it is invalid.
They merely remark that formalisms employing such actions risk superluminal
spectra. They presumably have a continuum formula (for speeds) in mind.
However, the validity of such formulae on the lattice cannot be assumed:
relativity modifies Newtonian formulae and quantum mechanics classical ones;
no continuum formula need survive latticization intact.
In any event, formalisms employing actions~\Ep{S.g} can, as a family,
simply not be any better or worse in their physical content than those
employing~\Ep{S.NN}: The two are related by a change of path variables.
There will be theories with good spectra and those with bad ones, but
neither's content can be affected by changing dummy variables.
[iv]~Most actions of the form~\Ep{S.g} will of course not describe neutrinos
or anything else specific, but that is irrelevant; all any system needs is
one~action.

The technical defect in the theorem's arguments, as given, is this:
they mingle parameteri\-z\-ation-independent assumptions and
parameteri\-z\-ation-dependent ones. They show that on a lattice the
following are incompatible:
[i]~A formalism describes neutrinos (chirality).
[ii]~It conserves probability (unitarity).
[iii]~It conserves momentum (translational invariance).
[iv]~Its action is {manifestly} local.
[v]~The action is {manifestly} hamiltonian.
However, only the first three of these are independent of our choice of the
dummy variables and thus genuinely characterize a \It{theory}; the last
two do depend on that choice and only characterize a~\It{parameterization}.

Failing to distinguish a theory from a representation would not matter
in a calculation of observables; those are representation-independent.
Nor would it matter in an existence argument;
if something exists in any representation, it exists.
\It{However, in a non-existence argument, it is fatal;} you end up proving
nothing more than the non-existence of the examined parameterization.

In other words, what the \NN no-go theorem says it proves and what its
mathematical content really amounts to are two different things.

I summarize the status of the theorem, including what we can salvage from it:
\itemize
\itm
{\sc What is shown}. If a unitary lattice
theory possesses a parameterization simultaneously local and hamiltonian,
it does not describe a chiral system with a conserved local quantized charge.
\itm
{\sc What it does not mean}. That has no implication for the existence
or the non-existence of lattice theories of neutrinos. It tells us that
such theories, if they exist, will not possess a parameterization which
is simultaneously local and hamiltonian.
\itm
{\sc What it does mean}. If local unitary lattice theories of neutrinos
do exist, their locality will be manifest only in a non-hamiltonian
parameterization. This amounts to a necessary but insufficient, indeed
weak, constraint on local representations.
\enditemize\noindent
Far from spelling the non-existence of lattice neutrinos, the nonlocality
of a particular parameterization is not even a credible obstacle: Every
theory possesses nonlocal representations. Our task in practice is
\It {always} only to find \It{one}~local representation.

%%-----------------------------------------------------------------------------
%
\newsec{5.~Some Perspective.}
A conclusion of lattice non-existence for {any} particle is always likely
to involve some misunderstanding somewhere, for it is equivalent to asserting
that an observation of those particles will \It{prove} that spacetime
{does not} possess a lattice structure {even at unobserved} scales.
\Ignore %-------------
\reference{gedanken}
Such ``gedanken-conflict'' with the Scientific Method suffices to reject
non-existence claims; the particles in question need not have been observed.
This refers to statements with observable content, e.g., ``Particle~X cannot
exist on a lattice'';
any caveat which has observational consequences, e.g., locality (finite-range
interaction), unitarity (conservation of probability), or translational
invariance (conservation of momentum), is irrelevant.
Those concerning calculational details, e.g., ``A lattice description of~X
is not possible without this many extra equations or that many extra terms'',
are neither forbidden nor particularly~limiting
\endreference.\space
\endIgnore %----------

Since many ``no-go'' investigations essentially show, either by direct
counting or by calculating some algebraic or topological invariant,
that the propagator possesses extra poles%
\reference{poles}
See, for example, Pelissetto's proof of the \NN theorem:
A. Pelissetto, \jnl'\AP'182'177'1988'
\endreference,\space
I emphasize that without understanding a specific formalism, you cannot
conclude that such poles \It{even affect} its observables:
Let $f$ be a smooth function such that $f(0)=1$ and $f(\pm\pi)=0$.
Let $g = \prod_{\al=0}^{3}f (\l{}k_\al)$; then $g=1$ in the continuum limit,
and $g=0$ at Brillouin Zone. Imagine a healthy lattice action. Multiply
it by $g^2$. The new action yields a propagator with many spurious poles,
which however give no energy-momentum relation and represent no particles,
let alone related species. More importantly, a formalism which generates a
factor of $g$ at vertices will nullify all effect of these poles. Someone
who merely counts the zeroes of the action, or otherwise looks at the
action alone, will misjudge this hypothetical formalism quite badly.

To appreciate why the hamiltonian and local structures can be incompatible
in the first place, notice that both are scarce among the representations
of a theory: in fact, their frequencies are $1/2^{\aleph_0}$ and
${\aleph_0}/2^{\aleph_0}$ respectively\cite{representationology}\cite{aleph}.
They are logically unrelated as well. Therefore, just as a randomly selected
real number is unlikely to be an integer, these two features will not be
compatible for \It{most} lattice theories.
% (simultaneously present in a representation)

The \NN theorem has encouraged the belief that species-doubling is a
fermionic problem possessing some special link with chirality.
% see J.~B.~Kogut, \jnl'\RMP'55'775'1983', p. 818 %%%
This belief is incorrect. The doubling only has consequences for chirality
when the afflicted system is fermionic. It can afflict systems obeying
Fermi, Bose, or an intermediate
statistics ($\ps_1 \ps_2 = e ^{i\th_{12}} \> \ps_2 \ps _1$).
It occurs in such systems as the (time-dependent) Schr\"odinger equation,
solitons, and gravity%
\reference{doubling}
The Schr\"odinger case is elementary. The doubling for solitons is discussed
by J.~Govaerts, J.~Mandula, and J.~Weyers, \jnl'\PL'112B'465'1982';
that for gravity by P.~Menotti and A.~Pelissetto, \jnl'\PRD'35'173'1981'
\endreference.\space
Most continuum theories would not admit ``naive'' discretization without
the doubling\cite{tobe}.
This is simply a defect of that discretizing algorithm---no more, no less.
Elsewhere I introduce an approach which can discretize any continuum theory
without introducing the~doubling\cite{l4}.

One straightforward lesson here is simply that \It{the hamiltonian form
is less practical a tool on the lattice than it is in the continuum.}
That merits attention, but it is no calamity: The surprise here is not
the breakdown of the hamiltonian method,
\It{but that it happens even in continuous time}. I do not suggest that
this is unimportant---it may mean that the degree to which isolating
dimensions has been useful in physics is a continuum peculiarity%
\reference{smearing}
The infinitely sharp transverse localization, needed to isolate dimensions,
is unnatural in a space where a fundamental length exists and must limit all
resolution. It was suggested in A.~K.~Trivedi, \jnl'\PL'B230'113'1989' that
a local discretization of the Dirac theory might require abandoning sharply
localized operations. This is \It{derived} for the more basic Weyl case
in \Ref{l4}
\endreference
---but it should not be mistaken for the non-existence of lattice neutrinos.
It tells us that one particular continuum tool will become less useful
after latticization.%

Contrary to a widespread impression, the \NN analysis does not throw any
light on the general question of chiral anomaly: It examines a specific
parameterization of a free theory. The anomaly requires an interaction
and, being a computed consequence of the theory,
is independent of parameterizations.

%%-----------------------------------------------------------------------------
%
\newsec{6.~Conclusion.}
The mathematical arguments presented in the \NN theorem fail to support
the ``no-go'' claims made in it.

A physical theory can be parameterized in many ways. A parameterization
can be local and it can possess a hamiltonian structure.
Those are two independent attributes.
Given sufficiently unsuitable field variables, both can be masked;
the question is, can we choose variables so cleverly that both are manifest
simultaneously?
For physically interesting continuum theories the answer has been yes,
but that is not a logical requirement and for a particular lattice
theory it may well be no (for most, it \It{is} no).
The given arguments merely find that this is indeed the case for lattice
neutrinos: at most one of those two attributes can be chosen to be manifest
in any one representation. This has nothing to do with the existence
or the non-existence of lattice neutrinos.

For a broad lesson, we stand reminded that fundamental investigations,
especially those ending in a non-existence conclusion, should be kept
parameterization-independent. Even when we must parameterize, we should
avoid constraints irrelevant to physics, e.g., how the action {looks}.

More narrowly, the problem of obtaining a local    % translation-invariant
unitary lattice formulation of neutrinos remains open.
I shall suggest one approach elsewhere\cite{l4}.

I emphasize that while the \NN theorem fails in its originally intended
role as a non-existence statement, it leaves us with a necessary constraint
on local parameterization of		% translation-invariant
unitary lattice theories of neutrinos.
It is, of~course, a well-known characteristic of deep ideas and insights
that they often turn out to be useful in unexpected and unintended ways.

The fairy tale I mentioned in the beginning had a happy ending. Eventually
that society learned that the existence of any curve had nothing to do with
how you parameterize the plane. They came to recognize that using new
variables also meant learning new tools such as nonlinear equations. In time
they learned to describe all kind of curves in cartesian coordinates,
including circles. Let us hope that the story of chiral lattice theory will
be an equally happy one.

\newsec{Acknowledgment.}
I am profoundly grateful to Professor H.B.~Nielsen for kindly and
generously sharing his perspective with me, in person and in a long
stimulating correspondence.

\Informal
\bigbreak\noindent
(\It{Informal Acknowledgment.} Before submitting this article to hep-lat, I
spent approximately one year trying to persuade Holger Nielsen that the \NN
no-go theorem needs reassessment, and that physics and young students entering
it will be best served if he and Prof. Ninomiya undertook it themselves.
Obviously I did not succeed. However, the discussion itself, which went from
minor matters to rather deep ones, was exhilarating. Nor is that the only
reason for my gratitude to Holger: I could have hardly survived one
particularly bitter winter in Copenhagen without the thermal clothing he gave
me.)
\endInformal

%%-----------------------------------------------------------------------------

\nobreak\bigskip\nobreak
\line {\hfil \hbox to 2.25in {\hrulefill} \hfil}
\bigbreak\ListReferences

%%-----------------------------------------------------------------------------

\pj % Informal, use \\ for explicit linebreak
If I have not seen further,\\
it is because giants are standing upon my shoulders.\\
\pjauthor ANONYMOUS \rm(\circa 1675)
\endpj

%%-----------------------------------------------------------------------------

\bye